\documentclass[12pt,preprint]{aastex}
\usepackage{amsmath}

\usepackage{epsf}

\shortauthors{Zucker et al.}
\shorttitle{Andromeda IX}

\begin{document}

\title{Andromeda IX: A New Dwarf Spheroidal Satellite of M31}

\author{Daniel B.\ Zucker\altaffilmark{1,10}, 
Alexei Y.\ Kniazev\altaffilmark{1}, Eric F.\ Bell\altaffilmark{1},   
David Mart\'{i}nez-Delgado\altaffilmark{1}, Eva K.\
Grebel\altaffilmark{1,2}, Hans-Walter Rix\altaffilmark{1}, Constance
M.\ Rockosi\altaffilmark{3}, Jon A.\ Holtzman\altaffilmark{4}, Rene
A.\ M.\ Walterbos\altaffilmark{4}, James Annis\altaffilmark{5}, Donald 
G. York\altaffilmark{6},
\v{Z}eljko Ivezi\'{c}\altaffilmark{7},  J.~Brinkmann\altaffilmark{8},
Howard Brewington\altaffilmark{8}, Michael Harvanek\altaffilmark{8},
Greg Hennessy\altaffilmark{9}, S.\ J.\ Kleinman\altaffilmark{8}, Jurek Krzesinski\altaffilmark{8}, Dan Long\altaffilmark{8}, Peter~R.~Newman\altaffilmark{8}, Atsuko Nitta\altaffilmark{8}, Stephanie A.\ Snedden\altaffilmark{8}}

\altaffiltext{1}{Max-Planck-Institut f\"ur Astronomie,
K\"onigstuhl 17, D-69117 Heidelberg, Germany; \texttt{zucker@mpia.de}}
\altaffiltext{2}{Astronomisches Institut, Universit\"{a}t Basel, Venusstrasse 7,CH-4102 Binningen, Switzerland}
\altaffiltext{3}{Astronomy Department, University of Washington, Box 351580,
Seattle WA 98195-1580}
\altaffiltext{4}{Department of Astronomy, New Mexico State University,
1320 Frenger Mall, Las Cruces NM 88003-8001}
\altaffiltext{5}{Fermi National Accelerator Laboratory, P.O. Box 500,
Batavia, IL 60510}
\altaffiltext{6}{Department of Astronomy and Astrophysics, University of Chicago, 5640 South Ellis Avenue, Chicago, IL 60637}
\altaffiltext{7}{Princeton University Observatory
Princeton, NJ 08544}
\altaffiltext{8}{Apache Point Observatory, P.O. Box 59, Sunspot, NM
88349}
\altaffiltext{9}{US Naval Observatory, 3450 Massachusetts Avenue, NW, Washington, DC 20392-5420}
\altaffiltext{10}{Guest Investigator of the UK Astronomy Data Centre}

\begin{abstract}

We report the discovery of a new dwarf spheroidal satellite of M31,
Andromeda IX,  based on resolved stellar photometry from the Sloan
Digital Sky Survey (SDSS).  Using both SDSS and public archival data
we have estimated its distance and other physical properties, and
compared these to the properties of a previously known dwarf
spheroidal companion, Andromeda V, also observed by SDSS.
Andromeda IX is the lowest surface brightness galaxy
found to date ($\mu_{V,0} \sim 26.8\, {\rm mag~arcsec}^{-2}$), and at the
distance we estimate from the position of the tip of Andromeda IX's
red giant branch, $(m - M)_0 \sim 24.5$ (805 kpc), Andromeda IX would
also be the
faintest galaxy known ($M_V  \sim -8.3$). 

\end{abstract}


\section{Introduction}
\label{txt:intro}

Hierarchical cold dark matter (CDM) models, while successful at large
scales,
predict too many low-mass dark subhalos to be
consistent with the observed abundance of dwarf galaxies, by at least 1 -- 2 orders of magnitude
\citep[e.g.,][]{klyp99,moor99,bens02b}.  
This problem can be at least qualitatively addressed if star formation in low mass
subsystems were inhibited, for example by photoionization
in the early universe; this could lead to galaxy luminosity functions with shallow faint-end slopes
at the present day \citep[e.g.,][]{some02,bens02a}, with observed
satellites embedded in much larger, more massive dark subhalos \citep{stoe02}.

Observational efforts to constrain the form of the
luminosity function for faint galaxies are hindered by the extremely low surface
brightnesses expected of such galaxies \citep[$\mu_{V} \gtrsim 26\, {\rm
mag~arcsec}^{-2}$, e.g.,][]{cald99,bens02b}. 
Conducting a comprehensive ground-based  survey for such diffuse
objects would be extremely difficult even in nearby galaxy groups.
Fortunately, in the case of the Local Group (LG), galaxies can be
resolved into stars, allowing one to reach much fainter limits
\citep[e.g.,][]{ferg02} and potentially place strong constraints on
both the LG luminosity function and galactic formation models.

In this letter, we report the discovery, using resolved stellar data from the Sloan Digital Sky
Survey (SDSS), of a new dwarf spheroidal
companion to M31, one which is the lowest luminosity, lowest surface brightness 
galaxy found to date. 
For this work, we have assumed a distance
to M31 of 760\,kpc \citep[$(m - M)_0 = 24.4$;][]{vand99}.

\section{Observations and Data Analysis}
\label{txt:obsdata}

SDSS \citep{york00} is an imaging and spectroscopic
survey that will eventually cover $\sim 1/4$
of the sky.
Drift-scan imaging in the five SDSS bandpasses ($u,g,r,i,z$)
\citep{fuku96,gunn98,hogg01} is processed through data reduction pipelines
to measure photometric and astrometric properties
\citep{lupt02, stou02, smit02, pier03} and
to identify targets for spectroscopic followup. For this work, we used
an SDSS scan along the major axis of M31, carried out on 5 October
2002 \citep[see][]{zuck04a} and processed with the same pipeline as Data Release 1 (DR1)
\citep{abaz03}.
In the
following, all references
to dereddening and conversion from the SDSS magnitudes to
$V,I$ magnitudes (for literature comparison) make use of \citet*{schl98} and
\citet{smit02}, respectively.

\begin{figure}[th]
\epsscale{1.0}
\plotone{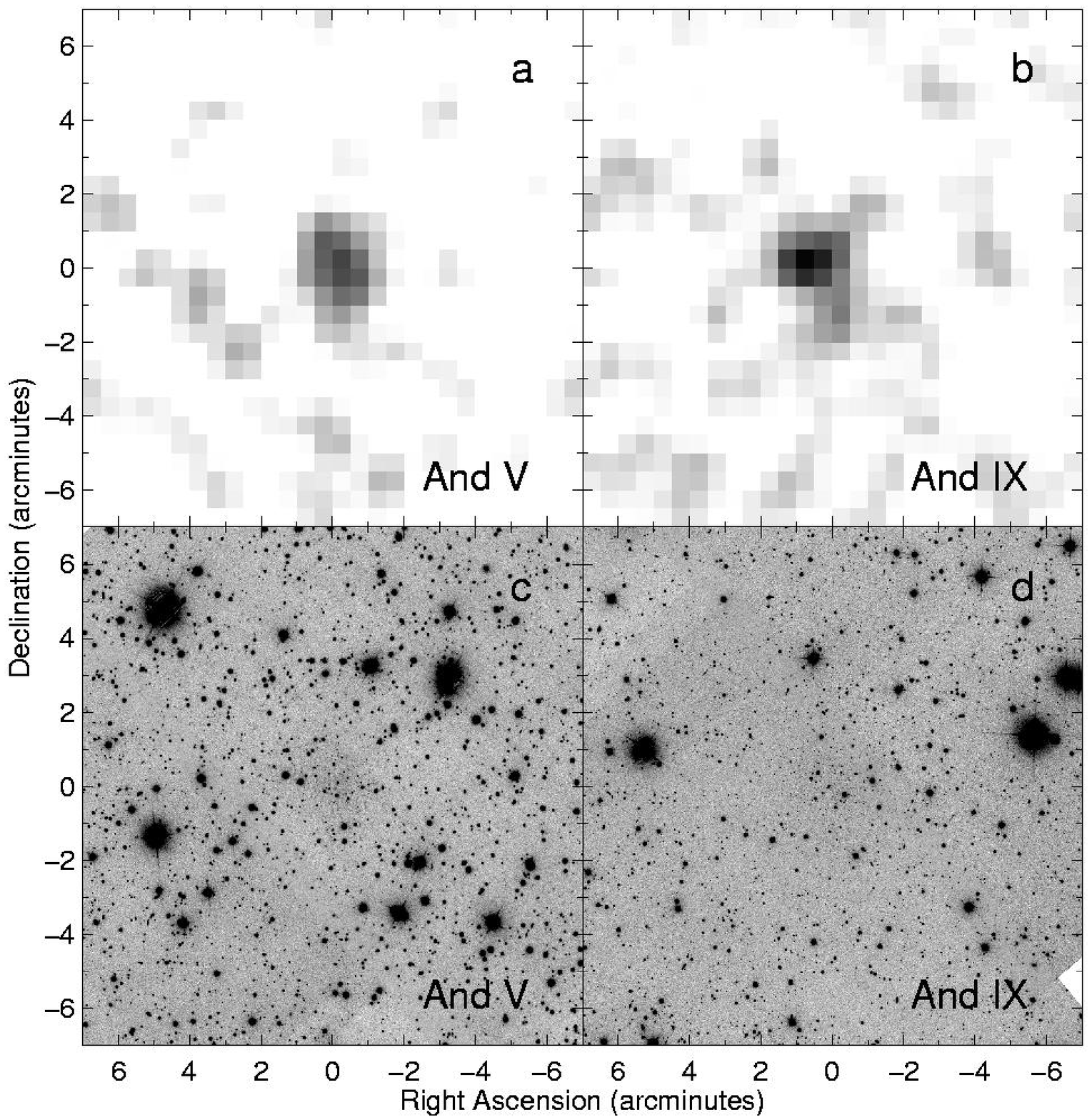}
\caption{\label{fig:sdssimgs}
And V and And IX as seen by SDSS:
{\it Top:} The spatial distribution of all SDSS stars with $i > 20.5$
in the fields of And V ({\it a}) and And IX ({\it b}). The data in both panels were binned $30\arcsec \times 30 \arcsec$ and smoothed with a Gaussian (FWHM$ = 1\arcmin$). 
{\it Bottom:} Combined $g$, $r$ and $i$ SDSS images of And V ({\it c})
and And IX ({\it d}). Each panel spans $14\arcmin \times 14 \arcmin$ and
is oriented with north up and east to the left.}
\end{figure}

\begin{figure}[th]
\epsscale{0.75}
\plotone{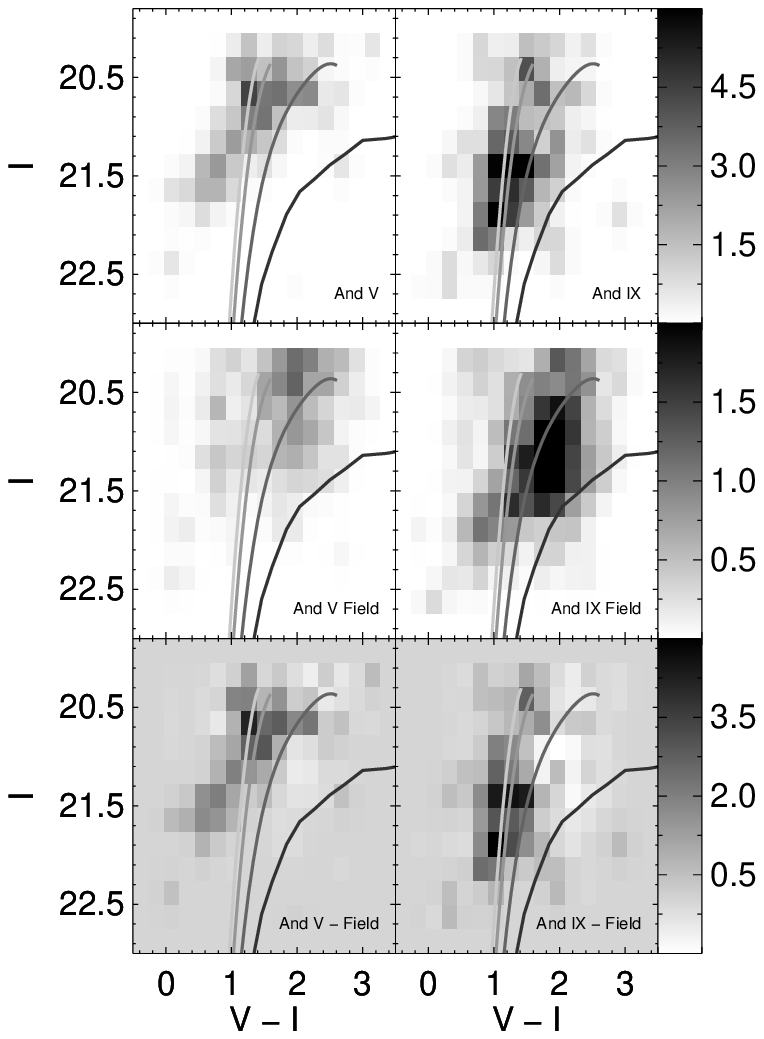}
\caption{\label{fig:sdsscmds}
Hess diagrams of And V ({\it left}) and And IX ({\it right}) from SDSS
Data: {\it Top}: 
Stars within a
$2\arcmin$ radius of the center of each galaxy. {\it Middle}: 
The control field for each galaxy (an annulus spanning
$8\arcmin$ to $10 \arcmin$ from each galaxy center). {\it Bottom}:
Hess diagrams for each galaxy, minus the control field Hess diagram (scaled
by the ratio of areas).
Data are dereddened, converted to $V,I$, binned by $0.25$ in $I$
and $V - I$, and smoothed with a Gaussian filter. The number of stars
in each bin is indicated by the grayscale to the right of each pair of
panels.
Fiducial sequences are overplotted
for Galactic globular clusters with metallicities of (left to right) [Fe/H] = $-2.2$ (M15),
$-1.6$ (M2), $-0.7$ (47~Tuc), and $-0.3$ (NGC~6553) \citep{daco90,
saga99}, shifted to the adopted M31 distance modulus of 24.4;  for
NGC~6553, a distance modulus of 13.7 and a reddening of $E(V - I) =
0.95$ were assumed \citep{saga99}. }
\end{figure}

This scan  
also included  
Andromeda V \citep[And V;][]{arma98}, a
known dwarf spheroidal (dSph) companion to M31;
applying a simple photometric filter to the SDSS data,
namely selecting all stars with $i > 20.5$, reveals the presence 
of And V (Fig.\ref{fig:sdssimgs}a), even though it 
is barely visible in the summed SDSS $g,r,i$ images of the field
(Fig.\ref{fig:sdssimgs}c). This same filter
yields another overdensity of faint stars (Fig.\ref{fig:sdssimgs}b) much closer to M31 on the sky, although it is not
apparent in the summed SDSS
image (Fig.\ref{fig:sdssimgs}d). In order to place some constraints on the nature of this stellar
overdensity, we compared the color-magnitude diagrams (CMDs) of stars in the fields of
And V and the new feature. For each object, we chose a circular region 
of radius $r = 2 \arcmin$, and 
annuli with $8\arcmin \leq r
\leq 10\arcmin$ as control fields for estimating foreground and field
star contamination. 
Fig. \ref{fig:sdsscmds} shows the resulting Hess
diagrams of And V and the stellar feature (top panels), the control
fields, scaled by the area ratio of target and control fields (middle
panels), and the difference of the two
(bottom panels).  The fact that
the red giant branch (RGB) in the new object extends to fainter magnitudes than that of
And V is explained by the significant differences in seeing ($\sim 1\farcs3$
vs. $\sim 1\farcs9$ in $i$-band)
and the consequent differences in completeness near the
faint 
detection limit.
RGB
fiducials for Galactic globular clusters spanning a wide range of
metallicities 
are overplotted.
The similarity between the two bottom
panels is striking, as both And V and the new stellar structure show
rather narrow, blue RGBs. While individual photometric errors and
uncertainties in the transformations between $g,r,i$ and $V,I$
photometric systems do not allow us to assign a specific metallicity
to either object, the blue color is significant; And V has a
metallicity of [Fe/H]$\lesssim -2$ \citep{davi02,guha00}, and the RGB color of the 
new feature suggests a comparably low metallicity.

\section{A New Companion of M31}

The stars in this new overdensity exhibit magnitudes and colors
consistent with their being near the tip of the RGB (TRGB), at approximately
the distance of M31 and its satellites. Their CMD is distinct from the 
surrounding field (cf. Fig. \ref{fig:sdsscmds}), and is in fact rather 
similar to that of And V, the known dSph satellite of M31 referred to
above; the similar angular size and extremely low surface brightness 
 are further indications that this structure is also a dSph galaxy. 
Following the 
nomenclature of \citet{vand72}, and in light of the recently proposed
Andromeda VIII \citep{morr03}, we have named the new object
Andromeda IX (And IX).

We applied the methodology of \citet{knia04} to the SDSS data; briefly, this entailed masking foreground stars and background
galaxies and subtracting a fitted sky level prior to measuring
integrated fluxes and central surface brightnesses.
The properties of And V derived in this way are shown in the first
column of Table\,\ref{tbl:pars}; \citet{cald99} found a central
surface brightness  $\mu_{0,V} = 24.8 \pm 0.2$ and an integrated
magnitude $V_{tot} = 15.42 \pm 0.14$ for And V, both in good agreement with our
results.
However, owing to And IX's extremely low surface brightness, no 
unresolved luminous component was detected in the SDSS
data. Consequently we used the data for And V (in which unresolved
luminosity could be quantified) to estimate corrections for And IX
surface brightness measurements based on resolved stars alone:
3\fm1$\pm$0\fm2, 2\fm6$\pm$0\fm2 and 2\fm3$\pm$0\fm2 for $g$, $r$ and
$i$ bands, respectively. These values are
consistent with those derived in \citet{zuck04a} from observations of
the Pegasus dwarf irregular and the Draco dSph.
Applying these corrections to the resolved-star surface brightness
profile calculated for And IX, we obtained dereddened central surface
brightnesses of 
27.0$\pm$0\fm5, 26.1$\pm$0\fm5 and 25.8$\pm$0\fm3
mag arcsec$^{-2}$ in $g,r,i$ (26.3$\pm$0.5 in  $V$, second column of Table\,\ref{tbl:pars}). Unfortunately, the small number of
stars detected in And IX (an excess of $\sim 60$ stars above the mean
field density in a $2\arcmin$-radius region) made it impossible to
reliably measure an integrated magnitude for And IX from the SDSS data.

\begin{figure}[th]
\epsscale{1.0}
\plotone{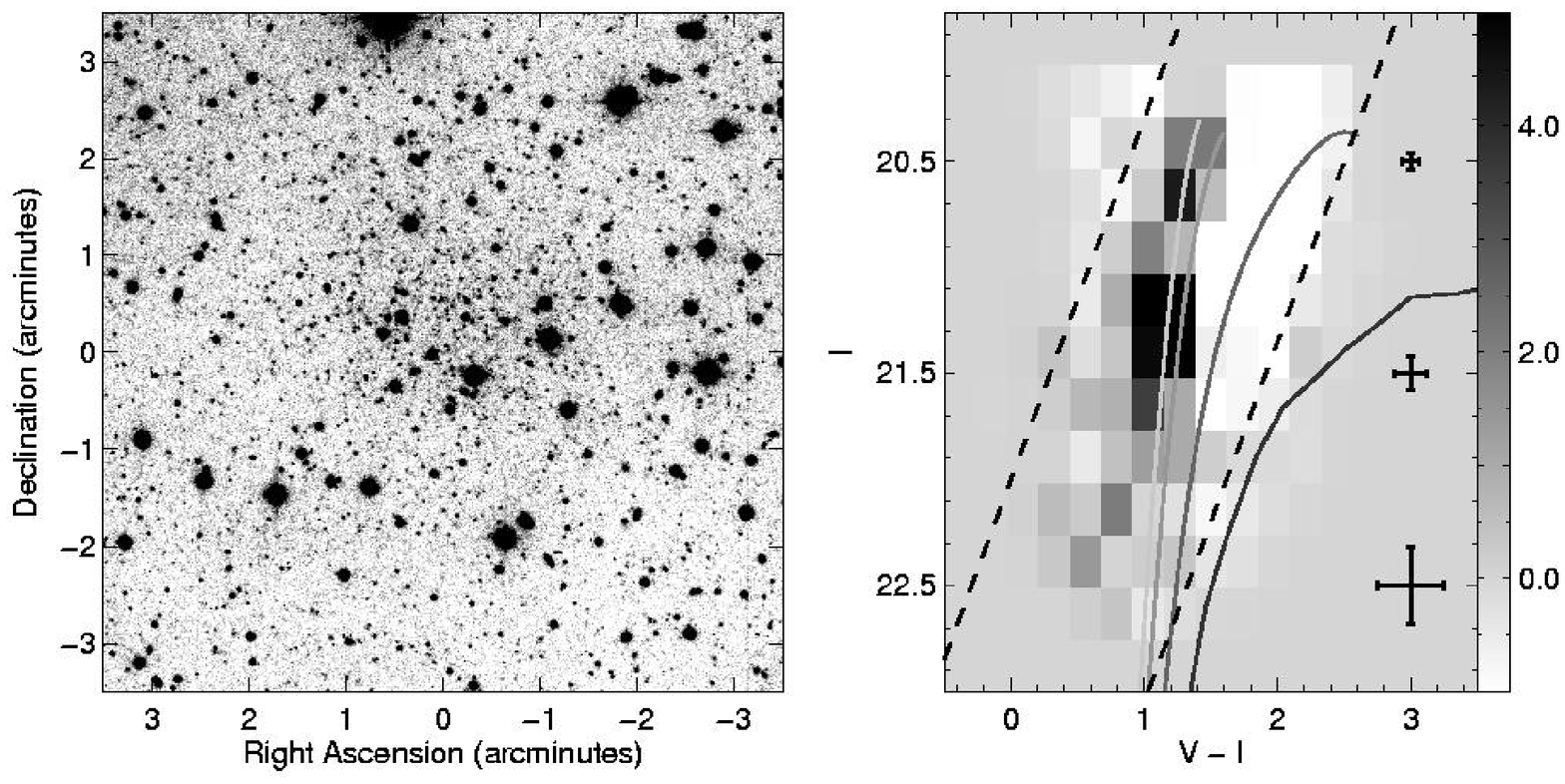}
\caption{\label{fig:intdata} And IX as seen by the INT: {\it Left}: An 800s
archival $V$-band image, covering $7\arcmin \times 7\arcmin$. {\it
  Right}: A dereddened, control-field-subtracted Hess diagram of And IX.
The dashed lines
indicate the limits of the color-magnitude region selected from both
SDSS and INT data for analysis of the $I$-band luminosity function
(see Fig. \ref{fig:lumfuncs}); the crosses show typical photometric
errors for stars over a range of $I$-band magnitudes.
Binning, smoothing, and the overplotted fiducials are the same as for
Fig. \ref{fig:sdsscmds}. The grayscale indicates the number of stars
in each bin.}
\end{figure}

\begin{figure}[th]
\epsscale{0.75}
\plotone{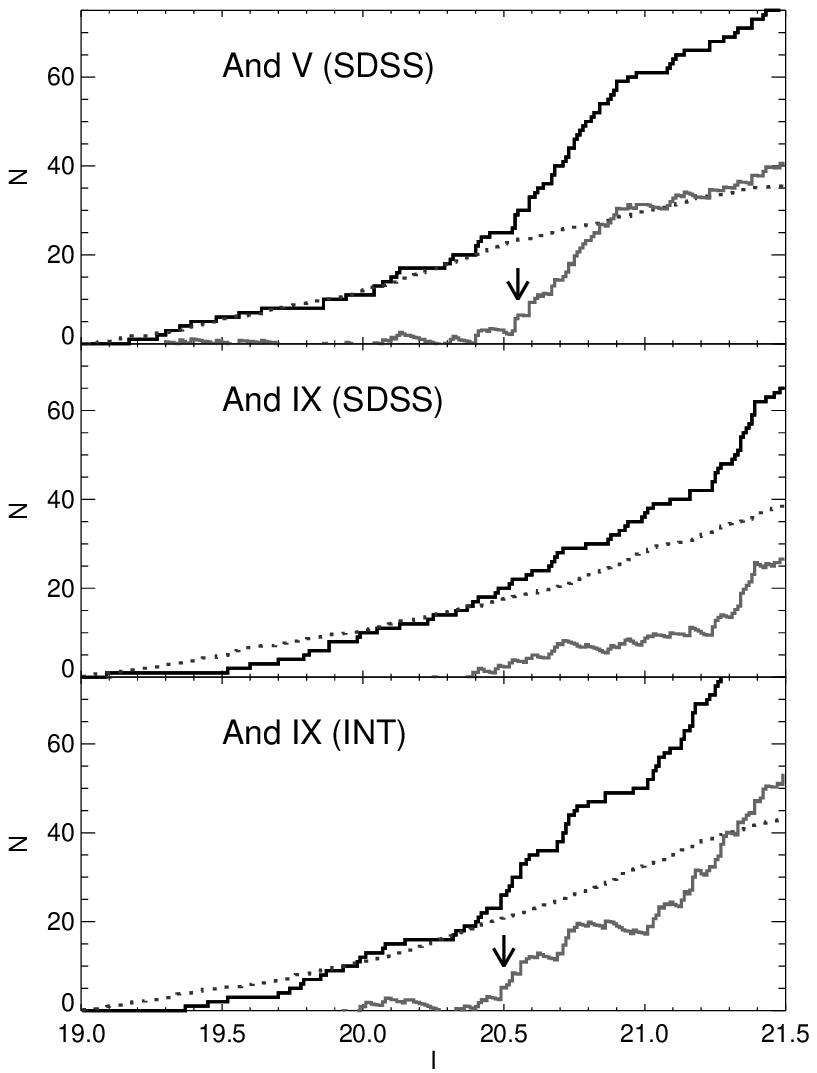}
\caption{\label{fig:lumfuncs} $I$-Band Luminosity Functions for And V
and And IX:
{\it Top}: Cumulative $I$-band luminosity function for SDSS stars within a $2\arcmin$ radius of the center of And V 
(black solid line), the scaled cumulative $I$-band luminosity function for the
corresponding control field (black dotted line), and the difference of 
the two (gray solid line). The arrow indicates the presumed location of the TRGB, $I = 20.55$.
{\it Middle}: The same as in the top panel, but for SDSS stars within
a $2\arcmin$ radius of the center of And IX.
{\it Bottom}: The same as in the top panel, but for INT stars within a
$2\arcmin$ radius of the center of And IX. The arrow indicates the presumed location of the TRGB, $I = 20.50$. All luminosity functions -- 
for both target and control fields -- are for dereddened stars falling in the
color-magnitude region illustrated in the lower panel of
Fig. \ref{fig:intdata}. 
}
\end{figure}

In order to learn more about And IX, we
analyzed
two images covering the region of the galaxy from the ING
Archive. These images, 800s exposures in $V$ and $i\arcmin$ filters,
were taken with the 2.5m Isaac Newton Telescope (INT)
on La Palma. The INT images
are significantly deeper than those from SDSS;
in fact, And IX is readily visible in the INT
$V$-band image (Fig. \ref{fig:intdata}, left panel).
We reduced the data using standard {\tt 
IRAF} tasks, obtained stellar photometry with DAOPhot \citep{stet94}, used our SDSS data to
bootstrap-calibrate this photometry to $V$ and $I$ magnitudes, and
dereddened the photometric data.

We then generated Hess diagrams for stars from the same
$2\arcmin$-radius region around the center of And IX as for the SDSS data, and from a nearby
$11\arcmin \times 11\arcmin$ control field, scaling the control field diagram by the area ratio of
target and control fields. The right panel of
Fig. \ref{fig:intdata} shows the 
difference of these two Hess
diagrams.
As in
Fig. \ref{fig:sdsscmds}, the control-field-subtracted Hess diagram for
And IX generated from the INT data reveals a narrow, blue RGB,
confirming that the observed stellar population of 
And IX is distinct from that of the projected field and
quite metal poor ([Fe/H] $\lesssim -2$).

The deeper INT images also provided better stellar number statistics and
allowed us to directly measure the unresolved luminosity of And
IX. Applying the same techniques as for the SDSS data, we determined the physical parameters listed in the third
column of Table~\ref{tbl:pars}. 
As a check on this method, we also tried psf-subtracting (rather than
masking) foreground
stars from the archival images before measuring physical parameters,
but the results differed by $\lesssim 0.05$\, mag from our initial
estimates (i.e., within the quoted measurement errors).
Of particular note are the dereddened
values for the central surface brightness, $\mu_{0,V} = 26.77 \pm
0.09$, and the integrated magnitude, $V_{tot} = 16.17 \pm 0.06$; not only do they
confirm And IX as the lowest surface brightness galaxy found to date,
but, at the distance of M31, 
And IX
would also be the lowest luminosity galaxy known.

The true distance to And IX is thus of considerable importance. One
method of measuring the distance to resolved, metal-poor galaxies like 
And IX is by determining the magnitude of the TRGB \citep[e.g.,][]{lee93}. 
We generated cumulative dereddened $I$-band luminosity functions for stars within the central
$2\arcmin$ radius of And V and And IX,
selecting only stars 
from the
color-magnitude region shown by the dashed lines in the right panel
of Fig. \ref{fig:intdata}. We did the same for the respective
control fields, scaling  
by the
ratio of target and control field areas, and then subtracted the
appropriate control field luminosity function from that of each
galaxy 
(Fig. \ref{fig:lumfuncs}). 
The field-star-subtracted luminosity
function begins to rise significantly 
at $I \sim 20.55\pm 0.15$ in And V and 
$I \sim 20.50\pm 0.15$ in And IX, which we interpret as the TRGB in
each galaxy.
Assuming metallicities of [Fe/H]$ =
-2.2$ and TRGB colors of $(V - I)_{TRGB} = 1.2$ for both galaxies, we
applied the empirical formulae of \citet{daco90} to 
derive a calibration for the TRGB distance: $(m - M)_0 =
I_{TRGB} + 3.98$. Thus we calculate And IX's distance modulus to be $\sim 
24.48$ (790 kpc), with errors on the order of $0\fm2\, (\sim \pm 70\,{\rm 
kpc})$. While this is a somewhat crude estimate, it is worth
noting that our distance to And V,
$(m - M)_0 \sim 
24.53$ (805 kpc) is in excellent agreement with that obtained by \citet{arma98},
$24.55$ (810 kpc).
The angular separation of And IX from the center of M31 is
$\sim 2\fdg6$, which at an M31 distance of 760 kpc
translates to a projected separation of 34 kpc; assuming that And IX
is some 30 kpc more distant than M31 places And IX at 45 kpc from the
center of M31. Thus, barring an improbably high relative velocity, And 
IX is a bound satellite of M31. 
At a distance modulus of $24.48$, the dereddened integrated magnitude
$V_{tot} = 16.17$ 
translates to an absolute magnitude of $M_V
\sim -8.3$, making And IX $0\fm6$ (a factor of 1.7) fainter than
the least luminous galaxy known, the Ursa Minor dSph \citep{kley98}. 

\begin{deluxetable}{lccc}
\tabletypesize{\small}
\tablecaption{Properties of And V and And IX  \label{tbl:pars}}
\tablewidth{0pt}
\tablehead{
\colhead{Parameter\tablenotemark{a}} & \colhead{And V} &
\colhead{And IX$_{\rm SDSS}$} & \colhead{And IX$_{\rm INT}$}
}
\startdata
R.A.(J2000.0)                    & ~~01 10 16.2        & ~~00 52 53.0    & ~~00 52
52.8       \\
DEC.(J2000.0)                    & $+$47 37 52         & $+$43 11 45     & $+$43 12
00        \\
A$_{\rm V}$           & 0\fm41              & 0\fm26          & 0\fm26
\\
$\mu_{\rm 0,V}$                  & $25\fm02 \pm 0\fm11$  & $26\fm3 \pm 
0.5$  & $26\fm77 \pm 0\fm09$ \\
V$_{\rm tot}$                   & $15\fm32 \pm 0\fm06$  & \nodata         &
$16\fm17 \pm 0\fm06$ \\
(m$-$M)$_0$                      & $24\fm53 \pm 0\fm20$  &   \nodata              &
$24\fm48 \pm 0\fm20$ \\
M$_{\rm tot,V}$                & $-9\fm2$          & \nodata
& $-8\fm3$ 
         
\enddata

\tablenotetext{a}{Surface brightnesses and integrated magnitudes are corrected for the mean
galactic foreground reddenings, A$_{\rm V}$, shown.}

\end{deluxetable}

\section{Discussion}

And IX is the lowest surface brightness,
lowest luminosity galaxy found to date, with a metallicity comparable
to the least chemically-evolved stellar systems in the LG; its
distance from the center of M31, $\sim 45$ kpc, places it well within, 
e.g.,
the virial radius (272 kpc) assumed by \citet{bens02b} in their
models, indicating that it is in all likelihood a satellite of 
that galaxy.
At an absolute magnitude of $M_V \sim -8.3$ ($\sim 2 \times 10^5$
L$_{\odot}$), And IX is comparable in luminosity to many globular
clusters, although two orders of magnitude larger ($r
\gtrsim 500 {\rm pc}$) .

Given that current hierarchical CDM models for galaxy formation still
generate a satellite galaxy luminosity function which rises at low luminosities
(albeit with a shallower slope than some earlier models), the discovery of And IX
raises some interesting questions. Is And IX a rarity, one of a small
number of such extremely low luminosity galaxies in the LG,
with the vast majority of the predicted large numbers of low mass
subhalos surviving only as dark matter? Or could And IX be the tip of
the iceberg, representative of a large population of low luminosity
dwarf satellites that have remained undetected
because of their extremely low surface brightnesses
\citep[e.g.,][]{bens02b}?

The discovery of And IX \citep[like that of And NE,][]{zuck04a} also highlights the capabilities of large-scale imaging 
surveys with uniform photometry, like SDSS, for detecting incredibly
diffuse stellar structures in the LG using their resolved
stellar components \citep[see][for a detailed discussion]{will02}. If
And IX is but one of a large population of extremely faint M31
satellites, it is quite 
likely that further analysis of the SDSS data
will yield more such objects, bringing the observed galaxy luminosity
function into better agreement with model predictions.

\acknowledgements

DBZ acknowledges support from a National Science Foundation International Postdoctoral Fellowship.
EFB acknowledges the financial support provided through the European
Community's Human Potential Program under contract
HPRN-CT-2002-00316, SISCO. This research has made use of data from the ING archive.

Funding for the creation and distribution of the SDSS Archive has been
provided by the Alfred P. Sloan Foundation, the Participating
Institutions, the National Aeronautics and Space Administration, the
National Science Foundation, the U.S. Department of Energy, the
Japanese Monbukagakusho, and the Max Planck Society.  The SDSS Web
site is (\texttt{http://www.sdss.org/}).

The SDSS is managed by the Astrophysical Research Consortium (ARC) for the
Participating Institutions. The Participating Institutions are The University
of Chicago, Fermilab, the Institute for Advanced Study, the Japan Participation
Group, The Johns Hopkins University, Los Alamos National Laboratory, the
Max-Planck-Institute for Astronomy (MPIA), the Max-Planck-Institute for
Astrophysics (MPA), New Mexico State University, University of Pittsburgh,
Princeton University, the United States Naval Observatory, and
the University of Washington.

\end{document}